\documentclass[prb,preprint]{revtex4-1}
\usepackage{amsmath}
\usepackage{amsfonts}
\usepackage{graphicx}
\usepackage{bm}

\begin{document}
\title[Readings of Gibbs]{Readings and Misreadings of \\ J. Willard Gibbs'\\ \emph{Elementary Principles in Statistical Mechanics}}

\author{George D. J. Phillies}
\affiliation{Department of Physics,Worcester Polytechnic Institute,Worcester, MA 01609}

\email{phillies@4liberty.net}

\date{\today}

\begin{abstract}

J. Willard Gibbs' \emph{Elementary Principles in Statistical Mechanics} was the definitive work of one of America's greatest physicists. Gibbs' book on statistical mechanics establishes the basic principles and fundamental results that have flowered into the modern field of statistical mechanics.  However, at a number of points, Gibbs' teachings on statistical mechanics diverge from positions on the canonical ensemble found in more recent works, at points where seemingly there should be agreement. The objective of this paper is to note some of these points, so that Gibbs' actual positions are not misrepresented to future generations of students.

\end{abstract}

\maketitle

\section{Introduction}

The objective of this paper is to clarify the textbook literature at points at which the textbook literature has diverged from Gibbs' \emph{Elementary Principles in Statistical Mechanics}\cite{gibbs1902a} at points where one might reasonably expect agreement rather than divergence. Some of these issues were previously noted in my textbook \emph{Elementary Lectures in Statistical Mechanics}\cite{phillies2000a}. I have here gathered these issues and some others. I do not claim originality in this approach.  In preparing this note, I shall be considerably less vigorous than was Gearhart\cite{gearhart1996a}, who enumerated a long series of texts that take different perspectives on one of these points, namely the specific heat of a diatomic gas.

Readers may legitimately ask how these interpretations crept into the textbook literature.  In some cases, the answer appears to be that many students of decades and decades ago learned statistical mechanics not by reading Gibbs, but by reading about his work as passed through the filters of Paul and Tatiana Ehrenfest and Richard C. Tolman. The Ehrenfests gave us the article \emph{The Conceptual Foundations of Statistical Mechanics}\cite{ehrenfest1990a}, which appeared in 1912 in the German \emph{Encyclopedia of Mathematical Sciences}.  The article was translated to English by M. J. Moravcsik in 1959\cite{ehrenfest1990a}.  Tolman actually gave us two textbooks on statistical mechanics, namely \emph{Statistical Mechanics with Applications to Physics and Chemistry}\cite{tolman1927a} and \emph{The Principles of Statistical Mechanics}\cite{tolman1938a}.  The latter is more widely read.  I note cases in which the interpretations that I am considering are already found in Ehrenfest or Tolman, but I am not claiming that these authors introduced rather than reproduced the interpretations in question. It is perhaps relevant that Ehrenfest was a student of Boltzmann, and approached statistical mechanics in considerable part from Boltzmann's perspective.

Readers interested in an extended development of alternative considerations on the foundations of statistical mechanics will find of particular interest Uffink's thorough review \emph{Compendium of the Foundations of Classical Statistical Mechanics}\cite{uffink2007a}. Uffink is heavily focused on the Boltzmann $H$ theorem and certain more recent issues, and only gives a short and partial treatment of the development in Gibbs' volume.

\section{Specific Heat of a Diatomic}

The issue that sparked my interest in misreadings of Gibbs was the specific heat of an ideal diatomic gas\cite{gearhart1996a}.  There is a standard textbook assertion that a classical diatomic molecule has three translational degrees of freedom and two rotational degrees of freedom. The rotations are around two axes that are perpendicular both to each other and to the line of centers between the molecules's two atoms.  The molecular kinetic energy is quadratic in the translational velocity and in the two rotation velocities, so from the equipartition theorem the average thermal energy of an isolated classical diatomic molecular should be $5 k_{B}T/2$. The corresponding specific heat is then $5 k_{B}/2$.

Gibbs had a completely different perspective on classical diatomic molecules.  He wrote (Gibbs\cite{gibbs1902a}, p.\ viii)

\begin{quote}
       ``\emph{...we do not escape difficulties in as simple a matter as the number of degrees of freedom of a diatomic gas.  It is well known that while theory would assign to the gas six degrees of freedom per molecule, in our experiments on specific heat we cannot account for more than five.}''
\end{quote}

To Gibbs, a diatomic molecule was expected to have  $6 k_{B}T/2$ as its average kinetic energy, not the $5 k_{B}T/2$ often assigned to it, and therefore correspondingly was expected to have $6 k_{B}/2$, not $ 5 k_{B}/2$,  as its specific heat. Six degrees of freedom? Yes, an extended rigid body has three translational degrees of freedom and three rotational degrees of freedom, a total of six degrees of freedom.  Some sources will propose that the moment of inertia $I_{3}$, for rotation around the axis parallel to the one bond of a diatomic molecule, is quite small, but it is a standard textbook exercise (Phillies\cite{phillies2000a}, Lecture 10, Problem 1) to show that the contribution of that rotation to the molecular specific heat is $k_{B}/2$, even in the limit $I_{3} \rightarrow 0$. It is sometimes worthwhile to point out to students that here we have a real physics calculation in which the $I_{3} \rightarrow 0$ limit of a calculated quantity is not equal to that quantity if $I_{3} = 0$ is assumed from the start of the calculation.

Once one has noted that Gibbs' actual positions are not always in accord with later authors' treatments of the same topic, one notices a series of other points at which Gibbs' analysis is not what might have been expected from the current pedagogical literature.  Here we consider the axiomatization of statistical mechanics, relationships between the canonical and microcanonical ensembles, the representative and canonical ensembles, identical particles, and the time dependence of the entropy.

\section{Axiomatization}

The most important topics for the description of a fundamental physical theory are the explicit axioms, implicit assumptions, and exemplary problems, models and results that describe the theory\cite{kuhn1970a}. This section considers the fundamental axioms of statistical mechanics, as treated by Boltzmann\cite{ehrenfest1990a}, Gibbs\cite{gibbs1902a}, and the representations of Gibbs' work by the Ehrenfests\cite{ehrenfest1990a}, Tolman\cite{tolman1927a,tolman1938a}, and more recent authors.

For Boltzmann, as described by the Ehrenfests\cite{ehrenfest1990a}, the fundamental axiom for the kinetic theory of gases was the \emph{assumption of equal a priority probabilities}, this being the assumption that all states of the system are equally likely to be found.

In contrast, Gibbs\cite{gibbs1902a}  (Gibbs, p.\ 33) asserted that the probability density for the most fundamental ensemble, this being the canonical ensemble, is
\begin{equation}
      P = \exp(-\frac{E - \Psi}{\Theta}).
     \label{eq:gibbs}
\end{equation}
Here $E$ is the energy of the given state of the ensemble. $\Theta$, the modulus, is eventually shown by Gibbs to have all of the properties of the temperature in natural units $k_{B}T$.  The normalizing factor $\Psi$ is eventually shown by Gibbs to have all the properties of the Helmholtz free energy $A$. According to Gibbs, the probability density $P$ is an exponential in the energy $E$. In Gibbsian statistical mechanics, all states of an ensemble do not have the same energy. Consequently, all states of the ensemble are not equally likely to be found.   Individual states are less likely to be found if their energy is larger.

In their treatment of Gibbs' book, the Ehrenfests\cite{ehrenfest1990a} refer to the problem of axiomatizing kinetic theory (Ehrenfest, p.\ 43) and propose that there is a ``\ldots certain instinctive knowledge\ldots'' to the effect that elementary occurrences ``...should in every instance be equally possible...”. They then claim that the microcanonical and canonical ensembles are with overwhelming likelihood the same. They offer a dubious path to reconciling their assumption of equal a priori probabilities with Gibbs' assumption of unequal a priori probabilities.  The Ehrenfests open by saying (Ehrenfest, p.\ 48)

\begin{quote}
“Taking into account the fact that the number of degrees of freedom in the gas model is the order of $10^{18}$, Gibbs proves the following theorem: In an ensemble which is canonically distributed with the modulus $\theta = \theta_{o}$, an overwhelming majority of the individuals will have very nearly the same total energy $E=E_{o}$.”
\end{quote}

They then say (Ehrenfest, p.\ 50)

\begin{quote}
``According to [GP: the previous quotation] it is first of all plausible that in general the average...over the canonical ensemble will be very nearly identical with the average value taken over a microcanonical or even ergodic ensemble with $E=E_{o}$...In a canonically distributed ensemble of gas models the overwhelming majority of the individual members are in a state given by the Maaxwell-Boltzmann distribution...with energy $E=E_{o}$. Thus, from the point of view of Boltzmann's presentation, the introduction of the canonical ensemble seems to be an analytical trick reminiscent of Dirichlet's integrating factor.''
\end{quote}

As the microcanonical ensemble follows the principle of equal a priori probabilities, and the canonical ensemble according to the Ehrenfests has with overwhelming likelihood the same distribution of states, it follows from the Ehrenfests' perspectives that the canonical ensemble's distribution follows, with overwhelming likelihood, the principle of equal a priori probabilities.

Their rationale has several gaps as a representation of Gibbs’ analysis:

First, Gibbs' book is not about the (ideal) gas model.  The book is about interacting particles.  For non-interacting particles of an ideal gas, the energy fluctuations are entirely the kinetic energy fluctuations. If the system is large, the kinetic energy fluctuations are indeed small relative to the average kinetic energy.  Gibbs does specify (Gibbs,  p.\ 73) for systems containing many interacting particles, that it is often the case that most elements of the ensemble will have very nearly the same energy. However, Gibbs then goes on to note (Gibbs, p.\ 75) that near a phase boundary, in the presence of an equilibrium between several phases, fluctuations in the potential energy and therefore the total energy are extremely large, so it is not true in statistical mechanics that energy fluctuations are always small. In their Footnote 179, the Ehrenfests refer to Gibbs' comments on his page 75, but only by remarking ``See the exception discussed in note 2, page 75 [of Gibbs]'' without noting precisely to what the exception is an exception. A reader of the Ehrenfests has no indication that Gibbs shows that there are commonplace conditions under which energy fluctuations are large, and therefore the Ehrenfest's rationale is incorrect.

Second, Gibbs viewed his canonical ensemble approach to be applicable even when the number of degrees of freedom of the system is small, perhaps only one or two.  Gibbs analyzes at some length   (Gibbs, pp.\ 171-176) the most probable values for the energy for three systems in equilibrium with each other, emphasizing the requirement that the number of degrees of freedom of each system must be greater than two (therefore, by inference, he considered that his theory could have been applied to a system having one or two degrees of freedom, but was not so applied here).  Discussing a feature of the three systems, whose number of degrees of freedom are $n_{1}$, $n_{2}$, and $n_{3}$, respectively, he remarks (Gibbs\cite{gibbs1902a}, p.\ 175):

\begin{quote}
    ``…These distinctions vanish for large values of $n_{1}$, $n_{2}$,  $n_{3}$. For small values of these numbers, they are important. Such facts seem to indicate that the consideration of the most probable division of energy among the parts of a system does not afford a convenient foundation for the study of thermodynamic analogies in the case of systems of a small number of degrees of freedom.''
\end{quote}

That is, Gibbs viewed as important and interesting that his treatment of statistical mechanics is valid if the number of degrees of freedom was small, for example, three. The Ehrenfests' rationale for the validity of Gibbs' statistical mechanics, namely that it agrees with Boltzmann, takes the opposite tack, namely the rationale requires that the system under consideration be large.

Tolman presents a description of statistical mechanics that is contradictory to Gibbs, while leaving innocent students with the impression that they are learning about Gibbs' model for statistical mechanics. Tolman (Tolman, pp.\ 60-61) introduces ``...the hypothesis of equal a priori probabilities for different regions of the phase space that correspond to extensions[GP: in phase space] of the same magnitude..." and

\begin{quote}
``...Thus, if we consider groups of neighboring states that correspond to differently located regions in the phase space $R_{1}$, $R_{2}$, $R_{3}$, etc. having the extensions in phase
\begin{equation}\label{eq:extension}
  v_{1} = \int_{R_{1}} \ldots \int dq_{1}\ldots dp_{f}, \ldots, {\rm etc.}
\end{equation}
we shall take the probabilities of finding the system in these different groups of states as proportional to the extensions $v_{1}$, $v_{2}$, $v_{3}$, etc., provided that our actual knowledge of the system is \textit{equally well} represented by the states in any of the groups considered. For example, if all we know about the system is that its energy lies in the range $E$ to $E+\delta E$, and we consider different extensions in the phase space $v_{1}$, $v_{2}$, $v_{3}$, etc. which themselves lie entirely in the range $E$ to $E+\delta E$, we shall take these extensions as giving the relative probabilities that the system itself would be found on examination to be in the corresponding different conditions. ... Under the circumstances we then have no justification for proceeding in any manner other than assigning equal probabilities to a system to be in different equal regions of the phase space that correspond, to the same degree, with what knowledge we do have as to the actual state of the system...''
\end{quote}

In applying eq.\ \ref{eq:extension}, Tolman\cite{tolman1938a} as quoted above gives \textit{as an example} a discussion of systems whose energy all lie in a narrow range $E$ to $E+\delta E$, $\delta E$ being small, this being the microcanonical ensemble.  However, he never restricts his \textit{principle of equal a priori probabilities} to systems having very nearly the same energy. If one instead took $\delta E>E$, but the regions still lay in the interval  $E$ to $E+\delta E$, the conditions claimed by Tolman for the applicability of the principle of equal a priori probabilities would still be applicable, but the principle of equal a priori probabilities would be entirely incorrect.  One reasonably infers that Tolman was so entirely focused on the microcanonical ensemble, in which $\delta E$ is small, that he did not find it necessary to note that his remarks were restricted to that ensemble and did not refer to Gibbs' canonical ensemble.

Tolman then claims (Tolman, p.\ 65) ``...the hypothesis chosen is the only postulate that can be introduced without proceeding in an arbitrary manner.'' However, Tolman makes this claim immediately  after introducing (Tolman, p.\ 58) the Gibbs canonical ensemble, which uses a contradictory postulate, namely that the a priori probabilities are unequal, so Tolman's claim that the principle of equal a priori probabilities is the only non-arbitrary postulate is contradicted by his own words. In Gibbs' canonical ensemble, the probability density as given by eq.\ \ref{eq:gibbs} is very definitely not a constant. Tolman attempts to reconcile these contradictory statements by claiming

\begin{quote}
      ``Under the circumstances ordinarily encountered in statistical mechanical applications, the canonical ensemble as defined by (22.4) [GP: i.e., eq.\ \ref{eq:gibbs}] is such that nearly all the systems in the ensemble have energies which are very close to the average energy for the ensemble.''
\end{quote}

Gibbs' canonical ensemble is thus claimed by Tolman to be practically the same as Tolman's microcanonical ensemble, this being the claim made previously be the Ehrenfests. However, Tolman (Tolman, p.\ 633) also describes (and correctly ascribes to Gibbs) systems readily encountered in the laboratory in which the range of energies represented significantly in the ensemble is large, namely systems in which there is an equilibrium between phases, e.g., solid and liquid, for example, a glass of ice water. Tolman claims that this large range of possible energies ``...again demonstrates the successful character of the statistical-mechanical explanation of thermodynamics...''.

Tolman's claim of a successful demonstration is logically self-contradictory. In his own earlier observations near his p.\ 60, Tolman notes (i) in the microcanonical ensemble, there are only infinitesimal energy fluctuations, and (ii) in Tolman's treatment of the canonical ensemble, the rationale showing that the canonical ensemble treatment is correct requires that the energy fluctuations are microscopically small. A calculation finding large energy fluctuations has now been obtained from a derivation that explicitly assumes that very nearly all states of the canonical ensemble have very nearly the same energy. The presence of large fluctuations in some systems vitiates Tolman's claim that the canonical ensemble follows from the principle of equal a priori probabilities and the presence of only minimal energy fluctuations.

In his review, Uffink is considerably more precise:

\begin{quote}
While Gibbs had not much more to say in recommendation of these three ensembles than their simplicity as candidates for representation of equilibrium, modern views often provide an additional story. First,the microcanonical ensemble is particularly singled out for describing as ensemble of systems in thermal isolation with a fixed energy $E$.
\end{quote}

The qualification "...with a fixed energy $E$" is correctly inserted, to distinguish from ensembles in thermal isolation in which different states of the ensemble have different energies, as seen in the states of the canonical ensemble. This qualification is sometimes omitted; see below.

Some texts treat these alternative axiomatizations well.  For example, Hill\cite{hill1956a} (Hill, p.\ 12) gives two choices of fundamental postulate, namely the postulate of Gibbs (eq.\ \ref{eq:gibbs}) and the principle of equal a priori probabilities, each correctly associated with its corresponding ensemble. Hill's treatment of fluctuations (Hill, pp.\ 97-98) is also precise, namely he notes cases in which fluctuations are not negligible, namely small systems and systems near a phase boundary or critical point, and measurable physical quantities (e.g., light scattering from concentration fluctuations) that are directly proportional to the size of the fluctuations, making the fluctuations directly observable.

There is also the extremely interesting variant derivation due to Friedman\cite{friedman1985a}, who proposes a restricted principle of equal a priori probabilities, namely that the probabilities of two states are only equal if the two states under consideration have \emph{equal} energies.   Friedman obtains the canonical ensemble by considering two well-defined systems $A$ and $B$ that are each in weak contact with a large heat bath $H$ but do not elsewise affect each other. The unequal relative likelihood of finding states having \emph{different} energies is then an outcome of Friedman's derivation, based on the assumption that the likelihood is only a function of the energy. Friedman's analysis does not constrain the size of the systems $A$ and $B$ or the relative size of fluctuations. Wannier\cite{wannier1966a} previously used a similar derivation.

On the other hand, one may readily find textbooks that invoke the principle of equal a priori probabilities, without noting the qualification that only states of more-or-less equal energy are being considered. Readers may always assume that the quoted authors knew perfectly well that they were only discussing the microcanonical ensemble, but students are reasonably entitled to believe what was written.

Huang\cite{huang1987a} (Huang, pp.\ 129-130) writes  ``When a macroscopic system is in thermodynamic equilibrium, its state is equally likely to be any state satisfying the macroscopic conditions of the system.'' Huang then claims that in thermodynamic equilibrium a system must be a member of the microcanonical ensemble.  The issue of fluctuations is dealt with via the assertion  ``If [GP: fluctuations are not small] there is no unique way to determine how the observed value of $f$ may be calculated. When [GP: fluctuations are not small], we should question the validity of statistical mechanics.'' Huang's (Huang, p.\ 146) assertion that fluctuations are always small ignores the diverging specific heat at a phase transition, and is limited to large systems. Gibbs, as noted above shows that statistical mechanics is entirely applicable to small systems and to systems near phase transitions.

McQuarrie\cite{mcquarrie1976a} (McQuarrie, p.\ 38) is even more explicit:

\begin{quote}

``Since the canonical ensemble has been isolated from its surroundings by thermal insulation, we can applied the principle of equal a priori probabilities to this isolated system. In the form we wish to use it here, the principle of equal a priori probabilities says that every possible state of the canonical ensemble, that is, every distribution of occupation numbers $a_{1}$, $a_{2}$, \ldots consistent with eqs.\ 2.1 and 2.2 is equally possible.''

\end{quote}

McQuarrie's eqn. 2.2 is a statement about the energy of the system, namely $\sum_{i} a_{i} E_{i} = \mathcal{E}$.  However, this statement is preceded with ``Because each of the systems of the canonical ensemble is not isolated but is at a fixed temperature, the energy of each system is not fixed at any set value. Thus we have to consider the entire spectrum of energy states for each member of the canonical ensemble.''. That is, in McQuarrie's treatment, each state of the system has an energy $\mathcal{E}$, his eq.\ 2.2, but, according to McQuarrie, all energies $\mathcal{E}$ are allowed, so McQuarrie's eq.\ 2.2 does not force every state of his canonical ensemble to have the same energy.  McQuarrie's treatment of fluctuations (McQuarrie, p.\ 59) includes two different positions on the observability of fluctuations: ``We shall see, however, that the probability of observing any value other than the mean value is extremely remote...In addition, there are several statistical thermodynamical theories of solutions and light scattering based on fluctuation theory...''

\section{Canonical and Microcanonical Ensembles}

A brief aside appears helpful.  In addition to systems found in a microcanonical or canonical ensemble, there is the separate notion of an \emph{isolated} system, a system that does not exchange energy with its surroundings. Systems in a microcanonical ensemble are isolated. They evolve in accord with their internal Hamiltonians, so the energy of each system remains constant.  As is sometimes not emphasized, each system in a Gibbsian canonical ensemble also evolves in accord with its own internal Hamiltonian (Gibbs, p.\ 4), so the energy of each system in Gibbs' canonical ensemble remains constant.  The systems in a canonical ensemble, as described by Gibbs, are all isolated systems. When one refers to energy fluctuations within the canonical ensemble, what is meant is that different states of the system can have different time-invariant energies. There is an average energy $\langle E \rangle$, but all systems in the ensemble may not have the same energy, so the fluctuation $\langle (E - \langle E \rangle)^{2} \rangle$, the variation in energy from system to system, may be non-zero.

There is a standard textbook demonstration that one can imagine generating a canonical ensemble as a small system in weak thermal contact with a much larger system, the combined energy of the two systems being a constant.  This demonstration can lead to the impression that in a canonical ensemble each small system is in continual contact with its large system, so that the energy of the small system fluctuates in time. This impression is entirely contradictory to the canonical ensemble as created by Gibbs, namely each state of a Gibbsian canonical ensemble evolves in accord with its own internal Hamiltonian: the energy of each state of a Gibbsian canonical ensemble is a constant. Indeed, Gibbs discusses time-dependent external forces, as would appear if the small and large system were in continual contact, but only to say that time-dependent external forces would knock the system out of equilibrium, and therefore that the states of the canonical ensemble are not subject to time-dependent external forces.

Closely related to the question of axiomatization is the description of the relationship between the canonical and microcanonical ensembles. In Gibbs' work, a canonical ensemble fills some volume of phase space. Microcanonical ensembles were described by Gibbs as surfaces or thin sheets within the canonical ensemble phase space. Each sheet includes all states of the canonical ensemble having their energies between two narrowly spaced bounds. How narrow a volume? Gibbs (Gibbs, p.\ 116) notes the possibility of representing the possible inclusion of systems within a sheet by a Gaussian statistical weight, the Gaussian being in the difference between the energy of the system and the single energy defining the sheet. The Gaussian is finally being taken to the limit of zero width and unit area.  Readers will recognize this Gaussian and its limit as a description of the Dirac delta function.  The canonical ensemble was therefore seen by Gibbs to be formed of an infinite number of microcanonical ensembles, each microcanonical ensemble being wrapped around its neighbors of lower and higher energy much like the layers of an onion. Because the states of Gibbs' canonical ensemble are isolated, each state of the canonical ensemble initially belongs to some microcanonical ensemble, and remains there as time goes on.

From the standpoint of a student, the passage from the canonical to the microcanonical ensemble and its statistical weight is straightforward. Each microcanonical ensemble consists of all states of the canonical ensemble in which the energy $E$ has some fixed value $\mathcal{E}$, so from eq.\ \ref{eq:gibbs} all states of a microcanonical ensemble have the same statistical weight.

As noted above, many recent authors take the microcanonical ensemble as being fundamental, with each canonical ensemble being formed as a very small system within a much larger parental microcanonical ensemble. In this image, the principle of equal a priori probabilities for the microcanonical ensemble leads to unequal a priori probabilities for the states of the canonical ensemble. While Gibbs does note late in his book (Gibbs, ch.\ 10, 14) the possibility of generating a canonical ensemble from a small fragment of a microcanonical ensemble, it would be totally alien to remainder of his book to claim that this is how he viewed his canonical ensembles.

One might readily present a list of other textbooks which take the microcanonical ensemble to be fundamental, and present what is nominally a derivation of the canonical ensemble from the microcanonical ensemble.  In defense of all these authors, in writing a textbook, one is often trying to present an old story in a way that students will perhaps understand more clearly than in the past, not studying whether or not the old and widely accepted story could be improved in its accuracy. Khinchin\cite{khinchin1949a}, pp.\ 110-114, obtains a canonical ensemble as a small part of a microcanonical ensemble, and then claims that the Gibbs statistical weight  ``can only be based on the law for the isolated system'', a result contrary to Gibbs. Plischke and Bergersen\cite{plischke1989a} pp.\ 28-32, obtain the canonical ensemble as a small part of a microcanonical ensemble; they show that the energy fluctuations are proportional to the specific heat, and claim that in the limit of a large system the energy fluctuations are therefore a vanishingly small part of the total energy, thus ignoring phase transitions and small systems. Tanaka\cite{tanaka2002a}, pp.\ 58-61, obtains the canonical ensemble by considering many copies of the system of interest and a single very large system. Toda, Kubo, and Saito\cite{toda1992a}, pp.\ 49-51, obtain the canonical ensemble as a small part of a larger microcanonical ensemble, but in their derivation make no claim as to the size of fluctuations or the required size of the canonical ensemble.

\section{Canonical and Representative Ensembles}

It is worthwhile to note that Tolman invoked not the Gibbsian canonical ensemble but instead the \emph{representative ensemble}. While Gibbs does introduce (Gibbs, page 5) the representative ensemble, if not by name, Gibbs clearly viewed the representative ensemble as a tool that might assist some students in envisioning the canonical ensemble. The two ensembles are not the same. Gibbs specifies:

\begin{quote}

In strictness, a finite number of systems cannot be distributed continuously in phase.  But by increasing indefinitely the number of systems, we may approximate to a continuous law of distribution, such as is here described.

\end{quote}

The distinction between the two ensembles will, alas, be less evident to students whose studies of mathematics stopped short of real analysis.

Let us consider a simple example that shows the distinction between Tolman's representative ensemble and Gibbs' canonical ensemble. The distinction corresponds to a practical question.  Suppose we wish to generate, experimentally, a series of numbers selected equally and approximately randomly from the integers 1, 2, 3, 4, 5, and 6. Such a series can be generated by rolling repeatedly a well-made six-sided die.

Now suppose we instead want to generate a random series of the same six integers, except that "6" is supposed to be found three times as often as the other five choices of integer. There are two ways to advance.  One way is to take an octahedral die, labelled "1" through "8" on its sides, and repaint the "7" and the "8" to read "6".  Each of the eight sides of the octahedron is equally likely to be rolled, but there are three times as many "6" sides as there are any of the other sides, so "6" is found three times as often.  The octahedron corresponds to the representative ensemble, in which some events are more probable than others because they are represented in the ensemble with more copies.

The second way to advance is to take a six-sided die to a good machine shop and carefully drill a small cylindrical hole, to the appropriate depth, starting at the "1" dot of the die. The boring is then refilled with a small plug made of lead or some other dense material. Finally, the outermost end of the boring is refilled and repainted so that, to an innocent observer, nothing has been done to the die.  This process, which may be illegal in some jurisdictions, is known as "weighting" a die. However, if the correct weighting has been performed, the die will roll "6" three times as often as it rolls any of its other five sides.   Each state of the system, each possible die roll, is present once and only once in the list of possible outcomes, but some die rolls are more likely than others. The weighted die provides an example of a canonical ensemble.

In Gibbs' canonical ensemble, the number density of phase space points is the same everywhere in phase space. Gibbs describes the phase space points as ``...are distributed in phase in some continuous manner...''. Integrals $\int dp dq$ over equal volumes of phase space therefore contain the same measure (``number'' is imprecise, because the quantity of phase space points is uncountably infinite) of points (Gibbs, p.\ 6). The relative importance of different phase space volumes in computing an ensemble average of a quantity is determined by the statistical weight $P$ assigned to each phase space volume, which gives the relative importance to the ensemble average of the points in each volume of phase space.

Tolman used the representative ensemble, in which $P$ functions as a Jacobian, namely $P(p,q)$ represents the actual number of phase space points in each phase-space volume $\int dp dq$, so that the total number of systems in the ensemble is said by Tolman\cite{tolman1938a} to be $N = \int dp dq P(p,q)$. The importance of different volumes in forming an ensemble average is then determined by how many phase space points are in each volume of phase space, with each phase space point of the representative ensemble being assigned by Tolman an equal statistical weight.

\section{Identical particles}

Suppose one has a system of particles which are absolutely identical with each other.  Under modern conditions one might think of an isotopically pure single chemical substance, an element or molecule.  Associated with such a system is the Gibbs ``paradox''.  There are actually two ``paradoxes'' related to the same problem, one in classical thermodynamics and the other in classical statistical mechanics. Gibbs appeared to view the thermodynamic problem as a point where something strange happened, not a true paradox.  In a true paradox, classical thermodynamics would have made two contradictory statements about the same system.  Gibbs clearly viewed the statistico-mechanical issue as a path to resolving an ambiguity in his theoretical model, and not a paradox at all.

The thermodynamic issue is fairly straightforward\cite{gibbs1875a,wiedeburg1894a}.  Suppose we have a container split in two by a removable partition. On each side of the partition is a volume of gas at some pressure $P$ and temperature $T$.  If the partition is removed, the two volumes of gas interdiffuse and mix.  If the two volumes held different gasses, say krypton and xenon, as a result of the interdiffusion there would be a non-zero change in the entropy of the system, an entropy of mixing.On the other hand, suppose the two volumes each held the same gas, at the same temperature and pressure.  in that case, if the partition were removed, molecules from each of the two volumes would sometimes pass into the other volume, but there would be no change in the entropy of the system. Gibbs then asked an interesting question.  What happens if the molecules on the two sides of the container start out being different, but are gradually made more and more like each other?  The process of making the molecules more and more similar to each other can be imagined to be a continuous process, but the transition from there being an entropy of mixing to there being no entropy of mixing is sharp.\ How can a continuous change in the properties of the molecules lead to a sudden change in the thermodynamic properties of the system? Within thermodynamics, there did not seem to be an answer.  (However, mathematical processes with this property are well known. Note the remark above on specific heats in which taking the $I_{3} \rightarrow 0$ limit and setting $I_{3}=0$ from the beginning lead to different answers.)

There is a corresponding statistico-mechanical question. An ensemble is a complete, non-repeating list of all the states of the system, together with the statistical weight to be associated with each state.  An ensemble average involves a simple sum over all the states of the ensemble, each state being included once and only once in the sum, with each state being assigned its correct statistical weight. Suppose, however, that two of the particles in the system are identical. What does it mean that the list is non-repeating?  What does it mean that each state is to be counted once and only once?

Gibbs postponed his discussion of systems containing identical particles until the final chapter of his book (Gibbs, ch. 15). He considered a system of $N$ particles having position coordinates ${\bm R}^{N} = ({\bm R}_{1}, {\bm R}_{2},\ldots, {\bm R}_{N})$ and corresponding momentum coordinates ${\bm p}^{N} = ({\bm p}_{1}, {\bm p}_{2},\ldots, {\bm p}_{N})$. A \emph{state} of the system is specified by listing the positions and momenta of all the particles, e.g., ${\bm R}_{1} = {\bm R}_{a}$, ${\bm R}_{2} = {\bm R}_{b}$, \ldots, ${\bm p}_{1} = {\bm p}_{a}$ , ${\bm p}_{2} = {\bm p}_{b}$, \ldots where ${\bm R}_{a}$, ${\bm R}_{b}$, \ldots, ${\bm p}_{a}$, ${\bm p}_{b}$, \ldots are numerical vectors. Note that what is in modern terms called the \emph{state} of the system Gibbs called the \emph{phase} of the system; correspondingly, what we  call a \emph{volume in phase space} Gibbs called an \emph{extension-in-phase}. Suppose each particle in the system is different from all the rest, so that particle 1 is a hydrogen atom, particle 2 is a helium atom, and so forth. If we exchange the positions and momenta of two particles, e.g., so that ${\bm R}_{1} = {\bm R}_{b}$, ${\bm R}_{2} = {\bm R}_{a}$, \ldots, ${\bm p}_{1} = {\bm p}_{b}$ , ${\bm p}_{2} = {\bm p}_{a}$, \ldots, we clearly have created a different state of the system.

However, what happens if the two particles are identical, so that there is no way to tell particles 1 and 2 apart?  If we exchange particles 1 and 2, do we create a new state of the system, or are we looking at the same state of the system that we had at the start?  In answering this question, it should be emphasized that in classical mechanics outside of statistical mechanics the question has no meaning. If we are calculating planetary orbits, and two of the planets are (admittedly a highly unlikely state of affairs) exact copies of each other, absolutely nothing happens if the two planets are interchanged. Indeed, there is no rational reason in classical mechanics to ask whether exchanging the two absolutely identical planets gives us a new solar system or not. Classical mechanics is only concerned with the evolution in time of the planetary positions. The time evolution is not affected by exchanging two identical planets.

The so-called \emph{Gibbs Paradox} of statistical mechanics corresponds to this question. Suppose we have a container having volume $V$, divided by a partition into two pieces each having volume $V/2$. The two volumes $V/2$ are filled with two different monoatomic gasses to the same pressure $P$. The entropy of each volume is readily calculated, up to a constant, using Gibbsian statistical mechanics. Now the partition is removed.  The two gasses interdiffuse and come to equilibrium. Gibbsian statistical mechanics is again used to calculate the entropy of the system.  Unsurprisingly, the entropy is found to have increased.  Now repeat the experiment, except the two volumes V/2 are filled with the same isotope of the same monoatomic gas. What happens if the entropy is calculated using Gibbsian statistical mechanics?  If one believes that exchanging two identical particles creates a new state of the system (we'll see in a moment how this assumption lurks in the calculation), then when the identical gas molecules interdiffuse the entropy of the system increases.  That increase, which is a figment of the calculation, is the so-called Gibbs paradox. I say so-called because Gibbs clearly believed that no paradox was present. Gibbs' actual position is found in the opening of Chapter 15 of his book

\begin{quote}
     If two phases [GP: in modern terminology, states] differ only in that certain entirely similar particles have changed places with one another, are they to be regarded as identical or different phases? If the particles are regarded as indistinguishable, it seems in accordance with the spirit of the statistical method to regard the phases as identical.
\end{quote}

The paradox is termed the  ``Gibbs Paradox'', leading to the impression that Gibbs himself believed that exchanging pairs of identical particles leads to new states of the system. The above quote makes clear that Gibbs believed for identical particles that the two sets of coordinates $({\bm R}_{1} = {\bm R}_{a}, {\bm R}_{2} = {\bm R}_{b})$ and $({\bm R}_{1} = {\bm R}_{b}, {\bm R}_{2} = {\bm R}_{a})$ refer to the \emph{same} state of the system. The multiple position integrals are not quite correctly set up, and generate the same state too many times.  Is this possible in classical mechanics?  Clearly the response of classical mechanics is \emph{nihil obstat!}  Nothing obstructs. Classical mechanics never asks whether or not two states are the same, so nothing in classical mechanics poses an obstacle to Gibbs' interpretation. Gibbs was clearly aware of the two alternatives. However, the canonical ensemble is a formal construction, with no internal reason to prefer one interpretation or the other, so Gibbs chose the alternative that is consistent with experiment.

As a further point of confusion, the Ehrenfests (Ehrenfest, p.\ 62) assert

\begin{quote}
In constructing the quantity $H$, Boltzmann from the beginning considered the question of whether all molecules are the same, i.e., (see sections 12b and d), whether they can be permuted among themselves, or whether one has a mixture of gases. This circumstance, however, is not considered at all in the definition of the quantity $\Sigma$. Consequently, Gibb's definition gives ``no information about the way the concentrations of the various types of molecules influence the additive constant and the expression for entropy.''
\end{quote}

The Ehrenfests cite Planck\cite{planck1904a} (Ehrenfest, p.\ 62)\cite{planck1904a} as having emphasized this point.  However, the claim is not consistent with Gibbs' book, which specifically discusses (Gibbs, ch.\ 15) entropies of mixtures. Instead, what happens in the book is that Gibbs presents a general development and then late in the book inserts his consideration of the question of identical particles. This is a standard pedagogical technique, in which a general issue is presented and elaborated, and only later are refinements on the general issue added. Readers who do not penetrate Gibbs' book to Chapter 15 would be unaware that Gibbs treated identical particles.

The classical implementation of Gibbs' position is a slight modification of the phase space integrals. For identical particles all in the same volume $V$, those integrals can be written
\begin{displaymath}
      \int_{V} d{\bm R}_{1}  \int_{V} d{\bm R}_{2} \ldots \int_{V} d{\bm R}_{N}  \int d{\bm p}_{1}  \int d{\bm p}_{2} \ldots \int d{\bm p}_{N}.
\end{displaymath}
Here $\int_{V} d{\bm R}_{i}$ denotes the integral of ${\bm R}_{i}$ over the volume $V$, while $\int d{\bm p}_{i}$ represents the integral of the momentum ${\bm p}_{i}$ of particle $i$ over all of its allowed values.  As written (except for sets of measure zero), the $N$ position integrals generate each combination of positions ${\bm R}_{a}, {\bm R}_{b}, \ldots, {\bm R}_{n}$ a total of $N!$ times. Gibbs compensated for this overcounting, relative to his opinion on identical particles, by specifying that for $N$ particles the phase space integrals needed to be divided by a factor of $N!$, because the phase space integrals generate each state of the system N! times. If the $N$ particles included several different species of identical particles, each present $\nu_{i}$ times, a product of factorials $\nu_{1}!  \nu_{2}! \ldots$ replaces the $N!$ as a divisor. With this division in place, the entropy of a system of identical particles or a mixture of them is properly behaved, namely the entropy is an extensive property of the system,  As Gibbs finds in his book, the so-called Gibbs paradox then does not arise.

As an aside, it is sometimes proposed that quantum mechanics explains and eliminates the Gibbs Paradox. In quantum mechanics of identical particles, the wavefunction is symmetrized, so that the $N!$ copies of the classical state, as generated by the classical partition function, are reduced to a single symmetrized state that is a sum of $N!$ terms. There is then only one state in the sum.  However, this proposal only hides the identical-particle difficulty, namely one can now ask whether in the ensemble average the symmetrized state should be counted once, or $N!$ times, once for each of its terms.

\section{Time Dependence of the Entropy}

Here we reach a fundamental difference between the Boltzmann and Gibbs models in their descriptions of the time dependence of a quantity that could be said to be the entropy.  Boltzmann, as described by the Ehrenfests, focused his conclusions on the $H$-theorem.  In this theorem, a quantity $H$ is defined.  $H$ is shown to have the properties one would expect for the negative of the entropy.  Boltzmann considered the time evolution of $H$ for an ideal gas. If $H$ was initially calculated for a  ``non-equilibrium'' distribution of gas molecules, as time went on $H$ would tend to decrease (i.e., $S$ would tend to increase) until equilibrium was reached, following which $H$ would be approximately stationary.

In Gibbs' treatment, an average over a canonical ensemble is an average over all possible states of the system. Gibbs does define a quantity $\eta$, which is much like Boltzmann's $H$, and which turns out to be the negative of the entropy $S$. Gibbs' $\eta$ is a constant of the motion, no matter whether it is calculated for the entire ensemble or for a fixed set of points as they move through phase space.

In Gibbs' treatment there are no non-equilibrium states. All states of the canonical ensemble are equilibrium states. The Ehrenfests protest, writing (Ehrenfest, p.\ 75)

\begin{quote}
On the other hand, it is also clear that Gibb's measure of entropy is unable to replace Boltzmann's measure of entropy in the treatment of irreversible phenomena in isolated systems, since it indiscriminately includes the initial nonequilibrium states with the final equilibrium.
\end{quote}

The Ehrenfest rationale is based on the claim that within an ensemble there are less-uniform states (``non-equilibrium states'') that with time relax into more-uniform (``equilibrium'') states and remain there.  The Gibbs perspective is that all states of the system are equilibrium states and are found in the equilibrium ensemble, though not all with the same probability. According to Gibbs, because the states of the canonical ensemble are isolated, the entropy of a canonical ensemble is time-independent. It is a constant of the motion.

To amplify on Gibbs: It is true that on the average a highly non-uniform state of the system will usually tend to evolve into a superficially more uniform state of the system.  However if one samples the ensemble at random, and watches the time evolution of the sampled states, the rate at which non-uniform states are observed to evolve into more uniform states must precisely match the rate at which uniform states are observed to evolve into non-uniform states. Readers will recognize this statement as the principle of detailed balance. We are at equilibrium.  The only way we can find a non-uniform state at some time in an equilibrium system is if the non-uniform state had evolved from a uniform state at some earlier time, an admittedly rare event.

The Ehrenfests' assertion also overlooks Gibb's treatment of irreversible processes in his chapter 13 (Gibbs, pp.\ 157-163). Gibbs treats reversible and irreversible processes, not as things that happen within an ensemble, but as arising from interactions between two distinct canonical ensembles. On treating processes as descriptions of interactions between pairs of ensembles, Gibbs shows that processes that appear to be reversible or irreversible indeed duplicate their expected thermodynamic behaviors.

\section{Discussion}

This brief note has considered a series of points at which the teachings of J. W. Gibbs on canonical ensembles do not always match what some more modern treatises represent to be the properties of canonical ensembles. We have considered the specific heat of diatomics, the axiomatic bases of statistical mechanics, the treatment of identical particles, applications of the method to small as well as large systems, and the time dependence of the entropy.

\end{document}